\documentclass[twocolumn,aps,showpacs,prb,epsf,graphics,psfig]{revtex4}
\usepackage{graphicx}
\usepackage{graphicx}
\usepackage{bbold}
\usepackage{color}
\usepackage[normalem]{ulem}
\usepackage{amssymb}

\begin{document}
\title{
A New Approach to Modeling the Microdosimetry of Proton Therapy Beams
}

\author{Ramin Abolfath$^{1,2\dagger}$, Yusuf Helo$^{3,4}$, David J. Carlson$^2$, Robert Stewart$^5$, David Grosshans$^1$, Radhe Mohan$^{1,*}$}
\affiliation{
$^1$Department of Radiation Physics and Oncology, University of Texas MD Anderson Cancer Center, Houston, TX, 75031, USA \\
$^2$Department of Radiation Oncology, University of Pennsylvania, Philadelphia, PA 19104, USA \\
$^3$Imanova Centre for Imaging Sciences, Invicro LLC, London, UK \\
$^4$Department of Medical Physics and Bioengineering, University College London, UK \\
$^5$School of Medicine, Department of Radiation Oncology, University of Washington, Seattle, WA 98195, USA
}

\date{\today}

\begin{abstract}
{\bf Introduction}:
To revisit the formulation of the mean chord length in microdosimetry and replace it by the particle mean free path appropriate for modelings in radio-biology.

{\bf Methods}:
We perform a collision-by-collision following by event-by-event Geant4 Monte Carlo simulation and calculate double-averaged stepping-length, $\langle\langle l \rangle\rangle$, for a range of target sizes from mm down to $\mu$m and depth in water.
We consider $\langle\langle l \rangle\rangle$ to represent the particle mean free path.

{\bf Results}:
We show that $\langle\langle l \rangle\rangle$ continuously drops as a function of depth and asymptotically saturates to a minimum value in low energies, where it exhibits a universal scaling behavior, independent of particle nominal beam energy. We correlate $\langle\langle l \rangle\rangle$ to linear density of DNA damage, complexities of initial lethal lesions and illustrate a relative difference between predictive RBEs in model calculations using mean chord length vs. the proposed mean free path.
We demonstrate consistency between rapid increase in RBE within and beyond the Bragg peak and $\langle\langle l \rangle\rangle$, a decreasing function of depth.

{\bf Discussion and conclusion}:
An interplay between localities in imparted energy at nano-meter scale and subsequent physio-chemical processes, causalities and pathways in DNA damage requires substitution of geometrical chord length of cell nuclei by mean-free path of proton and charged particles to account for a mean distance among sequential collisions in DNA materials.
To this ends, the event averaging over cell volume in the current microdosimetry formalism must be superseded by the collision averaging scored within the volume. The former, is fundamentally global attribute of the cell nuclei surfaces and boundaries and is characterized by their membrane diameters, hence such global indices are not appropriate to quantitatively represent the radio-biological strength of the particles and their RBE variabilities that is associated with the sensitivities to local structure of the collisions and their spatio-temporal collective patterns in DNA materials.
\end{abstract}
\maketitle
\section{Introduction}
ICRU Report 36 on microdosimetry~[\onlinecite{ICRU36}] provides general formulations for the mean chord length in terms of the geometrical dimension of the sensitive volumes of interest (SVOI).
Accordingly, $\langle l \rangle \equiv \overline{l}$ is the mean length of randomly oriented chords in that volume and can be calculated based on the geometrical structure of the target of interest. 
The mean chord length that is an intrinsic geometrical character of SVOI, results from the random interception of the SVOI by a straight geometrical line, superseded for a (single) physical track of a charged particle.
The chord length calculated as such is the mean of shortest Euclidean distances between pairs of end points, crossing closed surface of SVOI.

As originally introduced by Rossi, Kellerer and colleagues (e.g., see for example Refs.~[\onlinecite{Kellerer1971:RR,Kellerer_Rossi1972:CTRR,Kellerer1984:RR,Kellerer1985:Book,Kellerer1975:REB}]), the lineal energy, $y = \varepsilon / \langle l \rangle$,
is the energy imparted in a single event, $\varepsilon$, divided by the mean chord length, $\langle l \rangle$.
$y$ is a random variable analogous to particle linear energy transfer (LET). Accordingly, the randomness stems from $\varepsilon$ as the chord-length has been averaged out and it is a predetermined number.
For a convex volume, $V$, such as spherical geometries with area $A$, a pure geometrical calculation, first performed by Cauchy~[\onlinecite{Cauchy1908}], yields $\langle l \rangle = 4V/A$.
For a rectangular slab with the transversal thickness $d$ perpendicular to the unidirectional beam central axis, $\langle l \rangle = d$~[\onlinecite{Magrin2018:PMB}].
The mean chord length, as such, which is based on straight tracks in a micro-meter size in a SVOI, has been vastly used in microdosimetry literature as well as in modelings in the radio-biological effects of any field of radiation.

One of the first studies based on MC technique to calculate chord-length distributions was reported by Birkhoff {\em et al.}~[\onlinecite{Birkhoff1970:HP}].
In this type of MC calculations, particles are not scored via a collision-by-collision approach using coarse-grained quantum mechanical scattering cross-sections, as we perform in now-a-days MC models such as Geant4~[\onlinecite{Agostinelli2003:NIMA,Allison2006:TNS,Allison2016:NIMPRA}] and Geant4-DNA~[\onlinecite{Incerti2010:IJMSSC}] where the tracks, including primary and secondary particles, exhibit the inchoate distribution of energy transfers~[\onlinecite{Conte2012:NJP,Abolfath2011:JPC,Abolfath2013:PMB,Abolfath2016:MP,Abolfath2017:SR,Abolfath2019:EPJD}].
Instead, Birkhoff {\em et al.}~[\onlinecite{Birkhoff1970:HP}] considered straight geometrical lines intercepting a sensitive volume of energy-proportional devices such as spherical / cylindrical gas proportional counters.
The rationale behind these simplifications is based on uniform and continuous energy transfer to matter along the tracks of charged particles.
Moreover, the tracks considered in simulations performed by Birkhoff {\em et al.}, were assumed to be part of a uniform source beam of radiation.

The MC techniques as such,
are therefore analogous to the classical Metropolis MC algorithm proposed for calculation of the numerical value of $\pi$, or numerical multi-variable integrations~[\onlinecite{NumericalRecipes}].
By taking a large random sample, the resulting chord-length distribution approaches the correct geometrical distribution in a statistical manner.


This is seemingly a plausible approach, applicable to cavities of the instruments designed for measuring pulse height distributions in a field of radiation such as energy-proportional devices, as details in localities of the collisions, their spatial distribution and compactness within the cavity sensitive volume are irrelevant to overall observable electrical responses. The mathematical model including the calculation of the chord length,
as presented in Birkhoff {\em et al.}~[\onlinecite{Birkhoff1970:HP}] and recent modifications and refinements in Refs.~[\onlinecite{Bolst2017:PMB,Anderson2017:MP,Bolst2018:PMB}], attempted to introduce an alternative approach to the Cauchy path length, is consistent with the occurrence of the physical processes in micro-dosimetry devices, considering Rossi's formulation of lineal energy $y$.

The responses from biological systems, however, seems slightly different from the physical responses we expect to observe from the design of micro-dosimetry proportional counters.
For example, the nano-meter scale linear density of DNA double strand breaks (DSBs) induced by traversing of a track of charged particle in a cell nucleus that determines complexities in lethal pathways of the lesions, cell death and tissue late effects, is expected to be more sensitive to the compactness and linear distribution of the individual collisions, rather than the geometrical dimensions of the cell nuclei~[\onlinecite{Abolfath2019:EPJD}].
The former roots in nano-meter scale localities in physio-chemical phenomenon and effects occurring in DNA material in cell nuclei whereas the latter is global characters associated with the outer structure of the cell nuclei boundaries and their membranes, prone to macroscopic-scale volume effects.
More precisely, such model calculations, as also pointed out, e.g., in Ref.~[\onlinecite{Carlson2008:RR}], only provide an estimate of overall imparted energy within cell nucleus volume without offering an appropriate analysis in resolving nano-scale biologically relevant events, as the geometric chord length of a cell nucleus is in $\mu m$-scale, three orders of magnitude larger than nano-meter scale, the microscopic resolution of DNA-damage.

Moreover, as particles penetrate in tissue, they lose their kinetic energy, hence the compactness of collisions and the complexities in DNA damage are expected to rise. The geometrical chord-length as calculated by Cauchy and implemented in microdosimetry, is a fixed number, e.g., a factor proportional to the dimension of cell nucleus and does not vary as a function of depth.

\begin{figure}
\begin{center}
\includegraphics[width=1.0\linewidth]{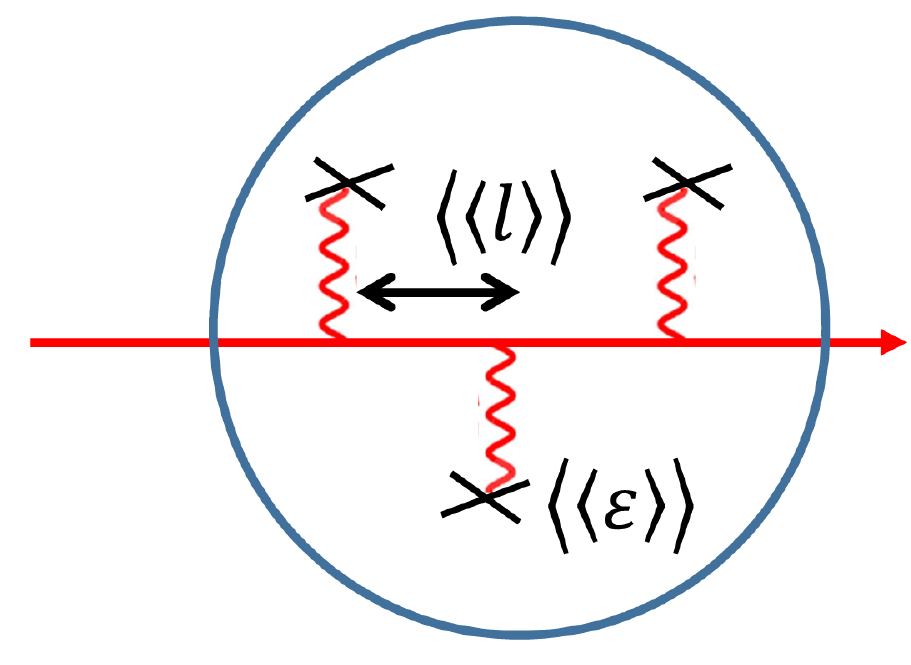}\\ 
\noindent
\caption{
Schematically shown passage of a charged particle (red bold arrow) through a $\mu$m-scale SVOI.
Collision-by-collision following by event-by-event length and imparted energy scales, $\langle\langle l \rangle\rangle$ and $\langle\langle \varepsilon \rangle\rangle$, respectively, describe typical distance between sequential interaction sites and imparted energy between charged particle and biological materials.
}
\label{fig0}
\end{center}\vspace{-0.5cm}
\end{figure}

Let us demonstrate the subtleties and lack of sufficiency in the standard microdosimetry formalism to fully describe and capturing the microscopic / nano-meter responses and critical variabilities in biological complexities, 
by considering passage of a single charged particle in a $\mu$m-scale SVOI, a typical representation of a cell nucleus, as schematically shown in Fig.~\ref{fig0}.
Hence occurrence of various damages in DNA-materials and chromosomes such as double strand breaks (DSBs) are expected to be seen in these volumes~[\onlinecite{Abolfath2011:JPC,Abolfath2013:PMB,Bianco2015:RRD}].

In Fig.~\ref{fig0}, the black cross-lines ($\times$) represent locations of the site of damages in DNA materials connected diagrammatically to the charged particle by the wiggly lines. The lines are Feynman diagram representation of the photon field propagators in quantum electrodynamics (QED) that describe interaction of charged particles in scattering processes with random interaction sites on DNA and/or the environment of DNA in a cell nucleus.
A process that describes release of OH-free radicals and/or reactive oxygen species (ROS) in indirect DNA damage (the latter), or shell electrons localized initially in DNA in direct damage processes (the former).
For further details in simulating such microscopic events at nano-scales, we refer the readers to our previous studies and publications, for example Ref.~[\onlinecite{Abolfath2013:PMB}].

For the present discussion, it is critical to recognize that the locations of the sites of interaction are random, hence the path lengths and distances between two sequential interaction sites along the beam central axis, $l$, are random variable. This is in addition to randomness in energy imparted at the location of interaction, $\varepsilon$.

Although it should be obvious that both $l$ and $\varepsilon$ are two random independent variables, following two independent distribution functions, in the standard microdosimetry formalism, only $\varepsilon$ was treated directly a random variable associated with the beam quality.
The sub-micrometer randomness in steps in $l$ was neglected, because such information was not necessary to be recorded from the measurement theory standpoints.
Instead a constant chord-length that is a characteristic / geometrical length of microdosimetric sites was superseded.

More rigourously, the statistical fluctuations in $l$ with regards to all collisions scored at nano-scale, crucial to DNA-damage statistical analysis, cannot be captured by micrometer-scale chord-length calculated by averaging $l$ over the track-distribution functions.
In fact, in the microdosimetry distribution functions, the location of the collisions over a single track were already traced out / integrated over thus any information regarding to individual collision was wiped out.
In our approach, presented in this work, we substituted the standard microdosimetry distribution functions by collision distribution functions.
Because a single track consists of several collisions in a microdosimetry site, the collision-based distribution functions of $\varepsilon$ and $l$ are more informative and appropriate for modeling biological responses.

In addition to aforementioned shortcomings, substantial number of studies and publications, including the microdosimetric kinetic model (MKM)~[\onlinecite{Hawkins1998:MP,Hawkins2003:RR,Kase2006}], implemented in the Particle and Heavy Ion Transport Code System
(PHITS)~[\onlinecite{Takada2018:JRR}], formulated based on an alternative approach by introducing a concept of virtual spherical domains in cell nuclei with a size that varies within nano- to $\mu$-meter.
In these models, it is not clear the rational for a choice of domain size but considering it spuriously an additional phenomenological parameter.
Therefore the application of current micro-dosimetry formulation in radio-biology does not seem appropriate and is prone to mixing the above discernible aspects.

We note that the Cauchy formula for the mean chord length is valid only for isotropic radiation (for non-spherical detectors)~[\onlinecite{Magrin2018:PMB}].
Hadron therapy beams are, at the contrary, non-isotropic and the primary particles have essentially a unique direction.
Thus several microdosimetry publications were developed based on non-Cauchy algorithms for the mean chord length.
However, the extension of Cauchy to non-Cauchy calculation of chord length does not seem to overcome the shortcomings of application of chord length in biological models, because the chord-length, by definition, is a character of a surface boundary of microdosimetry sites, whether it is a detector, cell nucleus or MKM domains.
The chord-length as such does not contain detailed information on occurrence of the biological events and complexities inside the bulk of the site.

Thus it is appealing to revisit and refine these well established micro-dosimetry models and generalize them for modelings in biological response theories, radio-biological applications, and nano-dosimetry.
This is the main goal of the present study to propose necessary modifications essential in tweaking micro-dosimetry and suit it for nano-dosimetry formulations.
Along this line of thoughts, we propose substitution of the cellular geometrical chord-length (regardless if it was computed using Cauchy or non-Cauchy techniques) by the tracks {\it diffusive mean free path}, $\langle\langle l \rangle\rangle$.

We demonstrate that to simulate biologically relevant passage of a charge particle, it is necessary to collect the events, collision-by-collision, and combine them together to shape a track-structure, a bottom-up approach.
This led us to perform a double averaging over collisions and tracks to calculate the track mean-free path, denoted by $\langle\langle l \rangle\rangle$.
Similar to a gas phase of matter, the molecules collide with one another, hence the diffusive mean free path is the average distance a particle travels between collisions. The larger the particles or the denser the gas, the more frequent the collisions are and the shorter the mean free path.
Turning back to original problem, the shorter the mean free path of the charged particle tracks manifests in higher complexities in DNA damage, hence the higher biological impact.
We thus supersede the track-averaged chord length, $\langle l \rangle$, by the mean-free path, $\langle\langle l \rangle\rangle$.
In calculation of lineal energy and LET, to be consistent with changes in calculating $\langle\langle l \rangle\rangle$, we also supersede the single event energy imparted, $\varepsilon$, by energy imparted in a single collision in the cellular SVOI. We then average over individual collisions and tracks in SVOI, wherever is necessary. 

We substantiate our proposal by presenting numerical illustrations on recently reported experimental data based on in-{\it vitro} clonogenic cell survival assay of non-small cell lung cancer (NSCLC) cells, i.e., an observation which was obtained by performing a high-throughput and high accuracy clonogenic cell-survival data acquired under exposure of the therapeutic scanned proton beams~[\onlinecite{Guan2015:SR}].
We compare two RBE models using (1) ICRU 36 geometrical chord length of typical spherical cell nuclei with radius of $\langle l \rangle = 5 \mu m$ and (2) charge particle diffusive mean-free path, $\langle\langle l \rangle\rangle$ that is a variable and a function of depth.
We show that a monotonic decrease in $\langle\langle l \rangle\rangle$ as a function of depth leads to a monotonic increase up to a factor of 4 in RBE.

\section{Method and Materials}
\subsection{Mathematical formalism}
In our recent study~[\onlinecite{Abolfath2019:EPJD}], we demonstrated that the linear density of DNA DSBs, $\Delta_l$,
a fundamental biological response function in a radiation field, is a variable of linear density of the collisions.
To this ends, we recall the following relation derived in our first principle multi-scale study~[\onlinecite{Abolfath2019:EPJD}]
\begin{eqnarray}
\Delta_l = \frac{\mu}{m}
\frac{\langle\langle \varepsilon^2 \rangle\rangle}{\langle\langle \varepsilon \rangle\rangle}
\frac{1}{\langle\langle l \rangle\rangle},
\label{eq1}
\end{eqnarray}
where $\varepsilon$ and $l$ are single collision energy imparted and stepping-length, calculated by Geant4 MC toolkit~[\onlinecite{Agostinelli2003:NIMA,Allison2006:TNS,Allison2016:NIMPRA}].
Here $\mu$ and $m$ are the average number of DSBs per deposition of 1 Gy of ionizing dose and cellular / DNA mass, respectively.

Hence, the relevant quantity in unit of lineal-energy (energy per length), consistent with Eq.(\ref{eq1}), was defined
\begin{eqnarray}
y_{1D} = \frac{\langle\langle \varepsilon^2 \rangle\rangle}{\langle\langle \varepsilon \rangle\rangle}
\frac{1}{\langle\langle l \rangle\rangle},
\label{eq2}
\end{eqnarray}
such that
\begin{eqnarray}
\Delta_l = \frac{\mu}{m} y_{1D}.
\label{eq3}
\end{eqnarray}

In terms of these quantities, calculation of cell survival was performed
\begin{eqnarray}
-\ln(SF) = \sum_{i=1}^N \sum_{j=1}^i b_{i,j} \Delta_l^{i-j} D^j.
\label{eq4}
\end{eqnarray}
Expanding Eq.(\ref{eq4}) and keeping the series up to quadratic term in dose, $D$, we found radio-biological $\alpha$, $\beta$ indices, as following
\begin{eqnarray}
\alpha = \sum_{i=1}^N b_{i,1} \Delta_l^{i-1},
\label{eq5_1}
\end{eqnarray}
and
\begin{eqnarray}
\beta = \sum_{i=1}^{N-1} b_{i+1,2} \Delta_l^{i-1},
\label{eq5_2}
\end{eqnarray}
The dependence of $\alpha$ and $\beta$ on $y_{1D}$ can be derived from the dependence of $\Delta_l$ on $y_{1D}$.
The coefficients in Eq.(\ref{eq4}) are identified by fitting to the experimental cell-survival data.

\subsection{Geant4 Monte Carlo simulations}
In Geant4 each pencil proton beams were simulated by irradiating a cylinder water phantom with 20 cm radius and 20 cm length. The mean deposited energies $\langle\langle\varepsilon\rangle\rangle$ and $\langle\langle\varepsilon^2\rangle\rangle$, the mean track length $\langle\langle l \rangle\rangle$ were scored within a linear array of voxels with 0.5 $mm$ down to 5 $\mu m$ thickness.
In our notations, averaging over all single point collisions, following by averaging over all single primary and secondary particle tracks are denoted by double averaging, $\langle\langle\cdots\rangle\rangle$.
In MC we scored $y_{1D}$, $y_{D}$ and LET$_d$ from collision-by-collision following by event-by-event averaging and energy deposition $\varepsilon_j$ and stepping length $l_j$, using the following identities
$y_{1D} = (\sum_j \varepsilon_j^2/\sum_j \varepsilon_j)/(\sum_j l_j/\sum_j 1_j)$, $y_D = \sum_j (\varepsilon_j/l_j)^2/\sum_j (\varepsilon_j/l_j)$, and
${\rm LET}_d = \sum_j (\varepsilon_j^2/l_j) / \sum_j \varepsilon_j$
where sum over $j$ includes all energy deposition events from primary and secondary processes in all steps and tracks in a specific voxel, hence $\sum_j 1_j$ represents total number of single-collisions, scored in all energy imparted events.

Therefore $y_{1D}=(\langle\langle\varepsilon^2\rangle\rangle/\langle\langle\varepsilon\rangle\rangle)/\langle\langle l \rangle\rangle$ as well as other types of LET's were calculated.
The number of primary protons and the number of interactions per track were saved, in the same volume.  Then, the energy deposition, the track length, the number of primary proton and the number of interactions were accumulated, in each cell. All simulations used $10^6$ protons with series of energy cut-off, corresponding to particle range that vary within 1 $mm$ and 1 $\mu m$.
In Geant4, any particle with energy below the cut-off value is assumed to not produce secondary particles (i.e. production threshold).
Below these cuts, the particle is transported further according to the CSDA approximation which will still imply a varying energy loss, i.e., no tracking cuts ~[\onlinecite{Agostinelli2003:NIMA}].
All simulation results presented used the QGSP-BIC-EMY physics list.
We used Gaussian proton energy spectrums with very small FWHM (0.18 MeV). Because of small divergence the simulated beam is mono-energetic.



In Geant4, when primary particle (in our case proton) collide with other particle, secondary particles are generated.
This includes photon, electron, proton, neutron, He and heavier ions. The primary particle will not be processed until all secondaries are dealt with.
The importance of the energy cut-off, requires the program to stop producing more secondaries when the energy of the secondaries become lower than energy cut-off, defined by the user as a cut-off in length-scale.
Otherwise the simulation will not be practical as will take considerably long time to simulate the events.
However, the very low energy secondaries will not be killed, but it will follow the continuous slowing down approximation, CSDA.
The CSDA works fine with low energy particles as they tend to not travel far in the material.
In other words, the cut-off represents the accuracy of the stopping position, and any particle will always be tracked down to zero kinetic energy.
In our previous studies, we did validate the Geant4 simulation using the cut-off values (as we used in current study) and experimentally measured Bragg peak via production of Cerenkov light~[\onlinecite{Helo2014:PMB}].
The difference was found to be less than 1\% for cut-off of 0.01 mm.

In the current manuscript, we used the terms cut-off, and cut-all interchangeably.
The term cut-all used in the legend of figures to emphasize a specific cut-off value applied uniformly for ``all" primary and secondary particles.
Note that the corresponding cut-off values in energy depends on the type of secondary particles.
For example the length-scale cut-off of 0.01 mm, approximately corresponds to 0.025 MeV cut-off for electrons in water.
Thus this value of energy for proton and the rest of particles is different than the corresponding value of cut-off in energy for electron.

Finally we note that the mean free path calculated by Geant4 methodology, as described above, should not be confused with the distance between individual ionizations, as this would be a true track structure approach. But from the Geant4 MC simulations, it must be clear what we calculated is a condensed history simulation where a large number of ionizations are grouped in a single energy loss step so it is rather the mean free path between individual energy loss events above a certain production cut.


\section{Results}

\begin{figure}
\begin{center}
\includegraphics[width=1.0\linewidth]{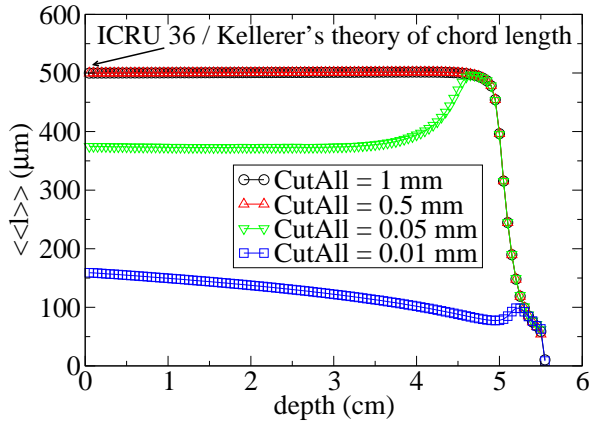}\\ 
\noindent
\caption{
Shown $\langle\langle l \rangle\rangle$ calculated for various cutoff values as a function depth for
a pencil beam of proton with nominal energy 80 MeV, traversing a cylindrical water phantom with slice thickness equal to 0.5 $mm$.
}
\label{fig1}
\end{center}\vspace{-0.5cm}
\end{figure}

In Fig.~\ref{fig1}, we illustrate the effect of particle cutoff length on $\langle\langle l \rangle\rangle$.
The calculation was performed for a pencil beam of proton with nominal energy 80 MeV, hitting a cylindrical water with radius and length of 20 cm, and slice thickness of 0.5 $mm$.
As shown in Fig.~\ref{fig1}, the depth dependence of $\langle\langle l \rangle\rangle$ is strongly influenced by the user defined cutoff choice.
The cutoff larger or equal to size of slice thickness, as shown by the black circles and red triangles respectively, yield the collision averaged chord-length equal to the slice thickness, a result that coincides with the geometrical chord-length of the water slice-thickness.
In this limit of large cut-off's, our numerical results recover the CSDA limits and the geometrical chord length calculated by Cauchy, Kellerer {\em et al.} and ICRU 36.
With lowering the cutoff down to 10th of slice thickness (green triangles) or even further (blue squares), the track mean-free path becomes significantly smaller than the geometrical slice thickness.
By lowering cutoff, below 0.01 mm, we observe negligible change in $\langle\langle l \rangle\rangle$,
i.e., if we lower the cut-off below 0.01 mm, the generated curve exhibits slight difference relative to the curve corresponding to 0.01 mm cut-off, shown by squares in Fig.~\ref{fig1}.
Numerically, we reached to a domain that resulted in convergence of the output data with respect to variations in cutoff values.
A continuous decrease in $\langle\langle l \rangle\rangle$ as a function of depth reveals an increase in linear compactness of collisions.

Note that, for slice thickness of 0.5 mm, as in Fig.~\ref{fig1}, and large cut-off values, no secondary particles are generated.
The primary particle, i.e., proton, loss energy according to CSDA and in results the simulated chord length turns out to be identical to the slice thickness, e.g., 0.5 mm as in Fig.~\ref{fig1}, which fits very well with Rossi-Kellerer's theory of chord-length.
However, if we lower the cut-off value the simulated chord length will become considerably less (0.15 mm as seen in Fig.~\ref{fig1}).

\begin{figure}
\begin{center}
\includegraphics[width=1.0\linewidth]{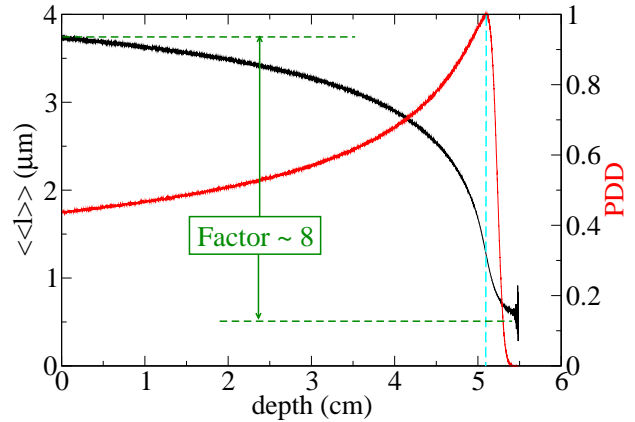}\\ 
\noindent
\caption{
Shown $\langle\langle l \rangle\rangle$ (black lines) and relative energy deposition (red lines) calculated for 80 MeV pencil beam of proton, traversing a cylindrical water phantom with 5.0 $\mu m$ slice thickness, using a cutoff values for all particles equal to 1 $\mu m$. The vertical dash line indicates the position of Bragg peak.
}
\label{fig2}
\end{center}\vspace{-0.5cm}
\end{figure}

In Fig.~\ref{fig2}, we show $\langle\langle l \rangle\rangle$ (black lines) and relative energy deposition, $\langle\langle \varepsilon \rangle\rangle$, (red lines) calculated for a pencil beam of $10^6$ protons with nominal energy 80 MeV, traversing a cylindrical water phantom with slice thickness equal to 5.0 $\mu m$.
To score such fine slices, we divided the water phantom thickness, equal to 20 cm, into 40,000 divisions and scored collision-by-collision energy deposition, $\varepsilon$, and stepping length, $l$, in each slice, using a cutoff values equal to 1 $\mu m$.
The vertical dash line indicates the position of Bragg peak at approximate 5.2 cm depth corresponding with $LET_d \approx 10 keV/\mu m$.

Note that because of 5 $\mu$m slice thickness and use of 1 $\mu$m cut-off, the effect of delta-rays and secondary particles, imparting their energies outside of the slices they were generated, effectively alter the shape of PDD. However, the location of the Bragg peak does not change as the slice thickness reduces from 1 mm to 5 $\mu$m.
We should also remark that if we lower the thickness of the SVOI to nano-scales, we must lower the cut-off values to below 1 nm to obtain sensible results.

It is intriguing to note that from beam entrance to the end of the proton range, $\langle\langle l \rangle\rangle$ decreases by a factor of eight.
Coincidentally, a factor of 4 has been reported experimentally in increase in proton RBE distal and proximal to Bragg peak (for more discussion see Fig.~\ref{fig4}, below).
The wiggling lines in the vicinity of proton range stem from statistical noises due to particle energy straggling and lack of enough statistics because of small number of particle fluence.

\begin{figure}
\begin{center}
\includegraphics[width=1.0\linewidth]{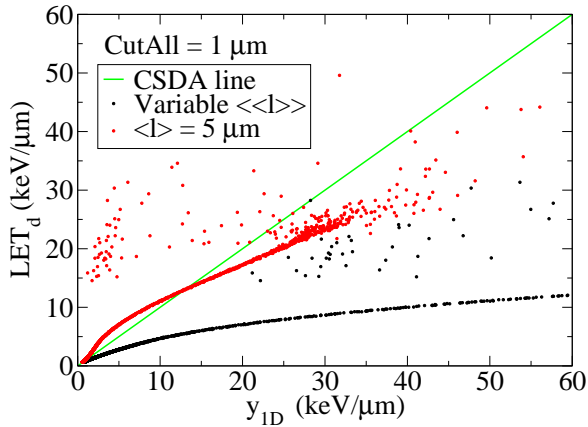}\\ 
\noindent
\caption{
Shown $y_{1D}$ vs. $LET_d$ scored for a MC setup identical to Fig.~\ref{fig2}.
}
\label{fig3}
\end{center}\vspace{-0.5cm}
\end{figure}

In Fig.~\ref{fig3}, $y_{1D} = \left[\langle\langle \varepsilon^2 \rangle\rangle / \langle\langle \varepsilon \rangle\rangle \right] / \langle\langle l \rangle\rangle$ vs. $LET_d = \langle\langle \varepsilon^2/l \rangle\rangle / \langle\langle \varepsilon \rangle\rangle$ is shown.
The trace shown by black dots obtained from variable $\langle\langle l \rangle\rangle$, as shown in Fig.~\ref{fig2} whereas the trace by the red dots uses an ICRU 36 type of chord-length estimate, i.e., $\langle\langle l \rangle\rangle = 5 \mu m$ everywhere, regardless of the depth.
Note that $\langle\langle l \rangle\rangle = 5 \mu m$ is a geometrical chord-length for a slab with thickness equal to $5 \mu m$.
The green solid line is a CSDA line, implying $LET_d = y_{1D}$ under assumption $\langle\langle 1/l \rangle\rangle \approx 1 / \langle\langle l \rangle\rangle$.
A similar non-linear dependencies between different types of LET's were previously presented in~[\onlinecite{Abolfath2019:EPJD}].
A non-linear dependence between LET$_d$ and $y_{1D}$ suggests advantageous in using $y_{1D}$ over LET$_d$ for RBE studies because of linear dependence between $y_{1D}$ and $\Delta_l$, as is given by Eq.(\ref{eq1}).

We note that the scattered dots in Fig.~\ref{fig3} are due to the statistical fluctuations in a MC simulation performed under limited number of primary protons. By increasing number of protons, the scattered dots merge together and form a single curve for each value of cutoff as shown in Fig.~\ref{fig3}.
In this calculation we used 10$^6$ protons. Because of small thickness of slices and small cut-off value, the MC calculation is very slow. To generate similar curves with lower statistical variations, a simulation based on larger number of protons, e.g., 10$^7$ protons is required. However, because in Fig.~\ref{fig3}, we only intend to show a non-linear dependence between LET$_d$ and y$_{1D}$, the envelope dependence obtained from scoring a million of protons is adequate for our current qualitative analysis. We note that, the red and black envelop-curves in Fig.~\ref{fig3} were collected through merging the scattered points. We also note that we did not perform any analytical fitting procedure to derive a specific curve, to investigate the analytical dependencies of LET$_d$ on $y_{1D}$.

\begin{figure}
\begin{center}
\includegraphics[width=1.0\linewidth]{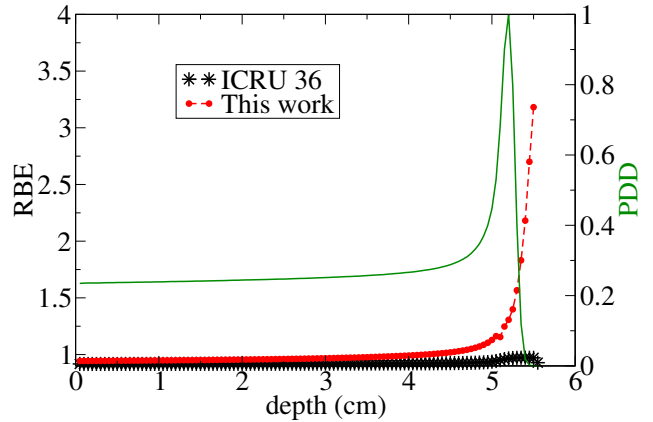}\\ 
\noindent
\caption{
Shown RBE calculated using variable $\langle\langle l \rangle\rangle$ (red dots) and constant chord length equal to $5\mu m$ (black dots), as recommended by ICRU 36.
RBE calculated after a surface fitting in 3D space to dose, LET and SF data for NSCLC H1437 cell lines.
A significant non-linear increase in RBE by a factor of $\approx 4$ in domains distal to the Bragg peak can be captured if $\langle\langle l \rangle\rangle$ is considered a variable of depth.
The biological endpoint used for calculation of RBE, corresponds to 10\% cell-survival fraction, as reported by Guan {\em et al.}~[\onlinecite{Guan2015:SR}].
}
\label{fig4}
\end{center}\vspace{-0.5cm}
\end{figure}

We now turn to present RBE of in-{\it vitro} clonogenic cell survival assay of H460 cell lines, a type of non-small cell lung cancer (NSCLC).
The dependence of SF on dose and LET were measured experimentally in our group~[\onlinecite{Guan2015:SR}]
and fitted by 3D global fitting method by the present authors~[\onlinecite{Abolfath2013:PMB,Abolfath2016:MP,Abolfath2017:SR,Abolfath2019:EPJD}].
Fig.~\ref{fig4} shows the result of RBE calculated using variable $\langle\langle l \rangle\rangle$ (red dots) and constant $\langle\langle l \rangle\rangle = 5\mu m$ (black dots),
corresponding a biological endpoint with 10\% cell-survival fraction, as reported by Guan {\em et al.}~[\onlinecite{Guan2015:SR}].
The numerical values of the coefficients in the polynomial expansion of $\alpha$ and $\beta$ on $\Delta_l$ and LET are given by Eqs.(\ref{eq5_1}-\ref{eq5_2}).
For $LET \leq 10 keV/\mu m$ we obtained the optimal fitting to the following polynomials with
$N=2$, $b_{1,1} = 0.05237$, $b_{2,1} = 0.0151$ and $b_{2,2} = 0.03265$, corresponding to reduced $\chi^2 = 0.00541$, $R^2$(COD)$=0.98233$ and adjusted $R^2 = 0.98233$.
This is a limit that dependence of $\alpha$ on $\Delta_l$ and LET is linear whereas $\beta$ shows no dependence on $\Delta_l$ (and LET),
Note that this is a limit frequently used in literature for all range of LET
(see e.g., Eq.(II.28) in Ref.~[\onlinecite{Hawkins1998:MP}] or Eq. (8) in Ref.~[\onlinecite{Kase2006}]).
For $LET > 10 keV/\mu m$, we obtained $N=6$ with
$b_{1,1} = 0.12288$, $b_{2,1} = 0.00624$,
$b_{3,1} = 1.322\times 10^{-6}$, $b_{4,1} = 9.449\times 10^{-7}$,
$b_{5,1} = 8.812\times 10^{-7}$,
$b_{6,1} = 7.725\times 10^{-8}$, and
$b_{2,2} = 0.00382$, $b_{3,2} = 0.0007821$, $b_{4,2} = 9.275\times 10^{-5}$, $b_{5,2} = 9.511\times 10^{-7}$,
$b_{6,2} = 1.68\times 10^{-7}$
corresponding to
reduced $\chi^2 = 0.05472$, $R^2$(COD)$=0.96777$ and adjusted $R^2 = 0.93911$.

The latter is a method of calculation, recommended by ICRU 36.
A comparison between two methods (a) considering variability in $\langle\langle l \rangle\rangle$ as a function of depth by double averaging over collisions and tracks and (b) considering geometrical chord length equal to the cell line thickness, $5\mu m$, evidently exhibits significant difference in predicting RBE by a factor of 4 in domains distal and proximal to the Bragg peak.

Similar variabilities in RBE as a function of proton range were reported in other experimental works using conventional microdosimetry approach and a methodology based on an empirical or phenomenological ``biological waiting functions" in calculation of RBE [\onlinecite{DeNardo2004:RPD}].
Accordingly, the biological waiting functions fitted to the spectrum of a spread-out Bragg peaks (SOBP) beam of proton collected by a 2.3 mm a microdosimetric prob, using a tissue-equivalent proportional counter (TEPC), resulted in a monotonic increase in RBE up to 2.5, a value close to the RBE reported in this work.
We note that the proposed modifications presented in this work, is free from such phenomenological convolution between the beam spectrum and RBE.

\section{Discussion and conclusion}
There are substantial evidence that proton (similar to heavier charged particles) RBE increases as a function of depth.
This is partly due to increase in spatial density of collisions as protons pass through tissue and lose energy at an increasing rate.
To make connection between experimental radiobiological data and mechanistic models, it is customary approach in literature to use microdosimetry models, developed by Rossi, Kellerer and colleagues~[\onlinecite{Kellerer1971:RR,Kellerer_Rossi1972:CTRR,Kellerer1984:RR,Kellerer1985:Book,Kellerer1975:REB}].
Embedded in these models, geometrical chord length of a SVOI plays a crucial role.
We discussed that although this quantity is a good parameter to describe microdosimetry processes and charge collections in tissue proportional chambers, but because it is intrinsically character of the surface and geometrical boundaries of cell nuclei, it cannot directly describe nano-meter scale localities in stochastic microscopic physio-chemical processes in DNA materials and DNA-damage.
Thus we proposed the diffusive mean-free path length of the particle tracks as a new metric that describes appropriately variabilities in collisions compactness as a function of depth in tissue to substitute the geometrical chord-lengths considered in micro-dosimetry modelings.
This variation of micro-dosimetry is potentially more appropriate for radio-biological studies.

{\bf Acknowledgement:}
The authors would like to acknowledge useful discussion and scientific exchanges with Drs. Alejandro Carabe-Fernandez and Alejandro Bertolet Reina.
The work at the University of Texas, MD Anderson Cancer Center was supported by the NIH / NCI under Grant No. U19 CA021239.

\noindent{\bf Authors contributions:}
RA: wrote the main manuscript, prepared figures, performed mathematical derivations and computational steps including Geant4 and Geant4-DNA Monte Carlo simulations and three dimensional surface fitting to the experimental data.
YH: contributed to Geant4 Monte Carlo simulations and writing the manuscript.
DC, RS, DG and RM: wrote the main manuscript, contributed to scientific problem and co-supervised the project.

\noindent{\bf Corresponding Authors:}\\
$^\dagger$ ramin1.abolfath@gmail.com / Ramin.Abolfath@pennmedicine.upenn.edu \\
$^*$ rmohan@mdanderson.org

\end{document}